\documentclass[conference, twocolumn]{IEEEtran}
\PassOptionsToPackage{table,xcdraw}{xcolor}
\usepackage{graphicx}
\usepackage{tikz}
\usetikzlibrary{positioning,fit,backgrounds,arrows.meta}

\usepackage{array}

\usepackage[utf8]{inputenc}
\usepackage[T1]{fontenc}

\usepackage{listings}
\usepackage{subcaption}
\usepackage[table,xcdraw]{xcolor}
\usepackage{amsmath}
\usepackage{bbm}
\usepackage{pifont}
\usepackage{amssymb}
\definecolor{mydarkblue}{rgb}{0,0.08,0.65}
\usepackage[colorlinks=true,
    linkcolor=mydarkblue,
    citecolor=mydarkblue,
    filecolor=mydarkblue,
    urlcolor=mydarkblue]{hyperref}
\usepackage{cleveref}
\usepackage{soul}
\usepackage[shortlabels]{enumitem}
\usepackage{url}
\usepackage{color}
\usepackage{tikz}
\usepackage{balance}
\usepackage{multirow}
\usepackage{comment}
\usepackage{wrapfig,booktabs}

\usepackage[symbol]{footmisc}
\usepackage{tablefootnote}
\usepackage{tabularx}
\newcolumntype{Y}{>{\raggedright\arraybackslash}X}
\usepackage{placeins}
\usepackage{algorithm}
\usepackage{algpseudocode}
\usepackage{ragged2e}
\usepackage{cuted}
\usepackage{float}
\usepackage{caption}

\usepackage{multicol}

\usepackage{dblfloatfix}
\usepackage{natbib}

\newcommand{\puffer}{PUFFER}
\newcommand{\lshbloom}{LSHBloom}
\newcommand{\fold}{FOLD}

\newcommand{\genW}[3]{%
  W\!\left(\begin{smallmatrix}#1 & #2\\[1pt] \varnothing\end{smallmatrix}\,;\,#3\right)}

\definecolor{codegreen}{rgb}{0,0.6,0}
\definecolor{codegray}{rgb}{0.5,0.5,0.5}
\definecolor{codepurple}{rgb}{0.58,0,0.82}
\definecolor{backcolour}{rgb}{0.95,0.95,0.92}

\makeatletter
\def\blfootnote{\xdef\@thefnmark{}\@footnotetext}
\makeatother

\lstdefinestyle{mystyle}{
  backgroundcolor=\color{backcolour},   commentstyle=\color{codegreen},
  keywordstyle=\color{magenta},
  numberstyle=\tiny\color{codegray},
  stringstyle=\color{codepurple},
  basicstyle=\ttfamily\footnotesize,
  breakatwhitespace=false,
  breaklines=true,
  captionpos=b,
  keepspaces=true,
  numbers=left,
  numbersep=5pt,
  showspaces=false,
  showstringspaces=false,
  showtabs=false,
  tabsize=2,
}

\AtBeginDocument{%
  \providecommand\BibTeX{{%
    \normalfont B\kern-0.5em{\scshape i\kern-0.25em b}\kern-0.8em\TeX}}}

\IEEEoverridecommandlockouts
\begin{document}

\title{
PUFFER: Incremental Fuzzy Deduplication for Continuously Evolving Corpora}

\newcommand{\corr}{\textsuperscript{*}}

\author{
Xiao Yang$^{\dagger,*}$, Erik Edward Aldape$^{\dagger,*}$, Beren Millidge$^{*}$
\\[0.5em]
\textbf{Zyphra}
\\
San Francisco, CA \\
\IEEEauthorblockA{$^{\dagger}$Equal contribution. \\ \textsuperscript{*}Corresponding authors: \texttt{\{xiao,erik.aldape,beren\}@zyphra.com}}
}

\maketitle

\setcounter{page}{1}


\begin{abstract}\normalfont\mdseries
Large language model training corpora are living collections that grow
through successive, often redundant releases, and each
release should be deduplicated against both itself and the accumulated history. At trillion-token scale, maintaining this history imposes a serious engineering challenge: the system must support incremental ingestion, bounded resident memory, deterministic retry, and dataset-scoped lifecycle control without repeated corpus-wide rebuilding. To address this, we constructed \puffer{} ({\bf P}rovenance-aware {\bf U}pdatable {\bf F}uzzy
{\bf F}iltering for {\bf E}volving {\bf R}epositories), a MinHash-LSH fuzzy-deduplication pipeline that
satisfies these constraints through two coupled system-level design choices. 
First, \puffer{} stores each LSH band as immutable, dataset-tagged, memory-mapped sorted segments, enabling exact historical band-key membership checks without RAM proportional to corpus size.
Second, $T$-fanout tiered compaction controls the number of historical segments searched during screening by periodically merging them, with a tunable cost in index-maintenance writes, and while preserving membership decisions.
Across $N$ ingested keys and $K$ equal-sized releases, \puffer{}'s cumulative maintenance cost is $\mathcal{O}\left(N\log N\log_T K\right)$ compared with $\Theta(KN)$ for repeated snapshot rebuilding. Dataset-tagged segments provide dataset-scoped lifecycle control: withdrawal is constant-time for datasets that remain uncompacted or protected, while withdrawal after compaction reconstructs only the affected merged segment rather than the corpus-wide index, even when the original dataset is no longer available.
Our implementation of \puffer{} completed the cumulative index-stage ingestion for one billion
documents in $\sim$1.75 hours in a single process, using 128 bytes
per document on disk for the $B{=}16$ band index. In comparison, a classical memory resident MinHash-LSH table required about 6.5\,KB per document in our baseline and was unable to finish in a configured 900\, GiB RAM cap. In a ten hour throughput comparison capped at one billion documents, \puffer{} achieved speedups of $11\times$ over \lshbloom{}, and $35\times$ over Milvus-LSH. 
\puffer{} is deployed in production on more than 30 billion documents, maintaining a continuously growing deduplicated corpus. We release \puffer{} as open-source software under a permissive license at \url{https://github.com/Zyphra/puffer}.

\end{abstract}

\section{Introduction}
\label{sec:intro}

Data curation is a central determinant of large language model quality. Modern open-corpus efforts have shown that training data quality depends on source selection, normalization, filtering, deduplication, domain balancing, and quality selection~\citep{
penedo2023refinedweb,
shen2023slimpajama,
soldaini2024dolma,
penedo2024fineweb,
li2024datacomp,
tokpanov2024zyda2,
weber2024redpajama,
penedo2025fineweb2}. Within this stack, fuzzy deduplication is especially
important. Exact deduplication removes byte-identical records; fuzzy deduplication instead targets
near-duplicates introduced by boilerplate, formatting changes, template reuse,
minor edits, or redistribution across sources. Such near-duplicates can waste compute, distort mixture composition, increase memorization risk,
and worsen train--test contamination
~\citep{lee2022dedup,kandpal2022dedup,carlini2022quantifying}.

Most existing data-curation pipelines are organized around building a snapshot
corpus where a large candidate pool is collected, filtered, deduplicated, and
ultimately used as a fixed training dataset. This workflow is increasingly mismatched
with how data is actually acquired. New Common Crawl dumps, code snapshots and domain-specific collections
continue to arrive after the previous corpus has been built, and datasets
occasionally must be removed due to poor quality, license takedown requests, compliance concerns or governance requirements. This means the corpus is not a static snapshot; rather it is a living collection that must be
maintained over time.
Existing alternatives are insufficient in several respects: they either require rebuilding snapshot state, rely on probabilistic capacity-provisioned membership structures, maintain resident approximate indexes, or lack support for dataset-scoped withdrawal. Appendix~\ref{app:related} provides a detailed comparison.

Our desiderata for an incremental deduplication pipeline designed for living corpora are as follows. First, the system must ingest new releases incrementally, so that each update touches only the incoming release and a
maintained historical index rather than rebuilding deduplication state over the
entire accumulated corpus. Second, it must remain faithful to the configured MinHash-LSH rule throughout its lifetime, requiring exact membership structures instead of probabilistic sketches whose approximation errors become part of the maintained state. Third, the historical index must be disk-resident and memory-mappable, so that querying accumulated history does not require a resident data structure whose RAM footprint grows with corpus size. Fourth, because failures are routine at production scale, ingestion must support deterministic retry. Finally, governance requires dataset provenance to be embedded directly into the index so that individual datasets can be inspected or withdrawn without rebuilding historical state.

\puffer{} is the only system, to our knowledge, that achieves these properties simultaneously. The
fuzzy-deduplication history is stored as immutable, dataset-tagged,
memory-mapped sorted segments of LSH band keys. Screening a new release is performed by
batched binary search over these segments. Committing a release atomically
replaces that dataset's tagged segment, which makes retries idempotent by
construction. Fixed-fanout tiered compaction merges segments using a streaming merge
executor with bounded working memory and acts as an explicit dial that
trades screening fanout against index-maintenance writes. Dataset-scoped withdrawal is a constant-time metadata operation while a
dataset remains uncompacted or protected; after compaction, withdrawal
requires reconstruction only of the affected merged segment rather than
the full historical index, without requiring access to the original dataset.

Our contribution is a practical, production-oriented index design for
continuously evolving deduplication corpora, combining a sorted-array layout
for band keys with decision-preserving compaction and dataset-level lifecycle
control, together with an analytical model of the resulting screening and
maintenance tradeoffs. This
lets compaction act as cost control rather than a change to deduplication
semantics; the full design is described in Section~\ref{sec:methods}. We show in Section~\ref{sec:results} that fixed-fanout compaction provides asymptotic advantages over naive snapshot rebuilding, and derive the optimal fanout under a steady-state cost model. We
implemented \puffer{} and evaluate its fidelity, scalability, resource efficiency, compaction tradeoffs, and dataset-lifecycle operations in Section~\ref{sec:results}.
\puffer{} completes the full index-stage lifecycle, from the input of band-keys to the output of the constructed deduplication index, for one billion documents
across forty releases in 1.75 hours in a single process. Compared with the closest implemented
alternatives, \puffer{} is the only design we evaluate that simultaneously
preserves exact MinHash-LSH band-key membership decisions, avoids a
corpus-size-proportional memory resident index, supports deterministic retry and
dataset-scoped withdrawal. Beyond index-level benchmarks, we demonstrate end-to-end operation on
2.5 billion documents (approximately 10 trillion tokens, Appendix~\ref{app:end_to_end}), from input Parquet
through deduplication and index construction to the final cleaned Parquet
corpus. Separately, \puffer{} has been deployed in production to process more
than 30 billion documents.

\section{Methods}
\label{sec:methods}

\begin{figure*}[t]
\centering
\begin{subfigure}[t]{0.56\textwidth}
\vspace{0pt}
\centering
\resizebox{\linewidth}{!}{%
\begin{tikzpicture}[
  font=\small,
  box/.style={draw, rounded corners, align=center, minimum width=3.0cm, minimum height=0.85cm, fill=gray!8},
  seg/.style={draw, rounded corners, align=center, minimum width=2.45cm, minimum height=0.65cm, fill=blue!6},
  op/.style={draw, rounded corners, align=center, minimum width=2.6cm, minimum height=0.75cm, fill=green!8},
  mini/.style={draw, rounded corners, align=center, minimum width=0.8cm, minimum height=0.45cm, fill=blue!6, font=\scriptsize},
  arrow/.style={-{Latex[length=2mm]}, thick}
]

\node[box] (incoming) {incoming dataset\\$D_k$, tag $k$};
\node[box, below=0.45cm of incoming] (keys) {band keys\\$v_b(d)$};
\node[op, below=0.45cm of keys] (screen) {(1) SCREEN\\binary search against\\all segments except tag $k$};
\node[op, below=0.9cm of screen] (commit) {(2) COMMIT\\atomically replace\\segment tag $k$};

\node[seg, right=1.1cm of incoming] (s1) {tag 1\\sorted keys};
\node[seg, below=0.25cm of s1] (s2) {tag 2\\sorted keys};
\node[font=\small, below=0.05cm of s2] (sdots) {$\vdots$};
\node[seg, below=0.05cm of sdots] (sj) {tag $j$\\sorted keys};
\node[seg, below=0.25cm of sj] (sk) {tag $k$\\this segment};
\node[draw, rounded corners, fit=(s1)(s2)(sj)(sk), inner sep=0.18cm, label={[align=center]above:historical index, band $b$}] (hist) {};

\node[op, right=1.3cm of s1] (compact) {(4) COMPACT\\$T$ same-level segments\\merge to level${+}1$};
\node[op, anchor=west, minimum width=2.6cm] (withdraw) at (compact.west |- sj) {(3) WITHDRAW\\drop distinct tag $j$};

\node[seg, minimum width=3.4cm] (merged) at ([yshift=-4.65cm, xshift=0.35cm]compact.south) {level-1 segment: union of\\$T$ level-0 tag segments};
\node[mini, anchor=south west] (m0a) at ([yshift=0.5cm]merged.north west) {L0};
\node[mini, right=0.14cm of m0a] (m0b) {L0};
\node[mini, right=0.14cm of m0b] (m0c) {L0};
\node[mini, right=0.14cm of m0c] (m0d) {L0};
\node[mini, right=0.3cm of merged, minimum width=1.0cm, dashed, fill=blue!14] (gone) {L1};
\node[font=\scriptsize, right=0.08cm of gone] (gdots) {$\cdots$};
\node[seg, minimum width=3.4cm, fill=blue!14] (ltwo) at ([yshift=-1.15cm, xshift=1.0cm]merged.south) {level-2 segment: union of\\$T$ level-1 segments};
\node[draw, dashed, rounded corners, fit=(m0a)(m0d)(merged)(gone)(gdots)(ltwo), inner sep=0.2cm,
  label={[font=\scriptsize, align=center]below:tier structure maintained by \textsc{compact} (example $T{=}4$)}] (tiers) {};
\node[box, below=0.8cm of commit, minimum width=3.3cm] (note) {after compaction,\\withdrawal requires\\sorted set-difference};

\draw[arrow] (incoming) -- (keys);
\draw[arrow] (keys) -- (screen);
\draw[arrow, dashed] (hist.west) -- node[midway, above, sloped, font=\scriptsize] {read} (screen.east);
\draw[arrow] (screen) -- node[midway, left, font=\scriptsize] {} (commit);
\draw[arrow] (commit.east) -| node[near start, below, font=\scriptsize] {atomic replace} (sk.south);
\draw[arrow] (s1.east) .. controls +(0.55,0.0) and +(-0.55,0.35) .. (compact.west);
\draw[arrow] (sj.east) -- (withdraw.west);
\draw[arrow] (s2.east) .. controls +(0.55,-0.2) and +(-0.55,0.1) .. (compact.west);
\draw[arrow] (compact.south east) -- ([xshift=-0.4cm]tiers.north east);
\draw[arrow] (m0a.south) -- ([xshift=-1.2cm]merged.north);
\draw[arrow] (m0b.south) -- ([xshift=-0.4cm]merged.north);
\draw[arrow] (m0c.south) -- ([xshift=0.4cm]merged.north);
\draw[arrow] (m0d.south) -- ([xshift=1.2cm]merged.north);
\draw[arrow] (merged.south) -- (ltwo.north);
\draw[arrow] (gone.south) -- (ltwo.north);
\draw[arrow] (tiers.west) -- (note.east);

\end{tikzpicture}%
}
\caption{Lifecycle of one band.}
\label{fig:index-diagram}
\end{subfigure}\hfill
\begin{subfigure}[t]{0.42\textwidth}
\vspace{0pt}
\footnotesize
\begin{algorithmic}[1]
\Require dataset $D_k$ with tag $k$; per-band segment sets $\{\mathcal{H}_b\}_{b=1}^{B}$; tier fanout $T$
\Ensure screened dataset $D'_k$; updated index state
\ForAll{$d\in D_k$} compute band keys $v_1(d),\dots,v_B(d)$
\EndFor
\For{$b=1,\dots,B$} group $D_k$ by $v_b$; keep the smallest stable identity per collision group forming $D'_k$ - save all of these band keys separately from the index
\EndFor
\For{$b=1,\dots,B$} \textsc{screen}: binary-search the current keys against every live segment of $\mathcal{H}_b$ except those tagged $k$
\EndFor
\State remove the union of within-dataset and historical matches, yielding $D''_k$
\For{$b=1,\dots,B$} \textsc{commit}: sort $U_{k,b}=\{v_b(d):d\in D'_k\}$; atomically \emph{replace} the segment tagged $k$
\EndFor
\For{$b=1,\dots,B$} \textsc{compact}:
\While{a level contains $T$ eligible segments} 
merge them
into one next-level segment, streaming as needed to respect
the RAM budget
\EndWhile
\EndFor
\State \textsc{withdraw} (on request, victim tag $j$): drop tag-$j$ segments while uncompacted or protected; otherwise rebuild the segment that contains tag-$j$ by taking the union of the band keys from each of the surviving datasets 
\end{algorithmic}
\caption{Incremental ingestion of one dataset $D_k$. Note: A segment is eligible for merging if it is unprotected and is not the most
recently committed segment.}
\label{fig:index-alg}
\end{subfigure}
\caption{Structure and lifecycle of the \puffer{} fuzzy index. (a)  Index layout for one LSH band (replicated across all $B$ bands), stored as immutable, dataset-tagged sorted segments.
(1) SCREEN: Binary-search keys against live historical segments. (2) COMMIT: Atomically replace segment for tag $k$. (3) WITHDRAW: Drop dataset tag metadata or rebuild affected segment. (4) COMPACT: Merge $T$ eligible same-level segments into tier $l+1$.   (b)~The per-dataset
ingestion procedure; step numbers (1)--(4) in diagram (a) mark where each
operation acts.}
\label{fig:index_structure}
\end{figure*}

We first formalize the incremental deduplication problem and then we describe how \puffer{} maintains
the corresponding band-key state across screening, commit, compaction,
retry, and withdrawal. Let $D_k$ denote the dataset release currently
being processed, let $k$ be its deterministic dataset tag, and let the live
history be the union of all non-withdrawn prior releases. Each release is
deduplicated against itself and against the live historical index before the
surviving documents $D'_k$ are committed. The MinHash-LSH configuration is
treated as fixed throughout an ingestion run.

\puffer{} uses the standard MinHash-LSH~\citep{broder1997resemblance} as the fixed fuzzy-duplicate rule that fixes our semantic notion of identifying near duplicates. We will now review this approach. Each document is represented as a set of
overlapping shingles, and two documents are considered similar according to
the Jaccard similarity of their shingle sets,
\[
    J(A,B)=\frac{|A\cap B|}{|A\cup B|}.
\]
Directly comparing all document pairs is infeasible at corpus scale. MinHash
compresses each shingle set into a length-$p$ signature
\[
    M(d)=[m_1(d),\ldots,m_p(d)],
\]
where agreement at each position estimates Jaccard similarity. Locality-sensitive hashing then partitions
the signature into $B$ bands of $R=p/B$ rows and hashes each band into a
64-bit key,
\[
    v_b(d)
    =
    \operatorname{xxHash64}
    \!\left(
    \operatorname{pack}
    [m_{(b-1)R+1}(d),\ldots,m_{bR}(d)]
    \right).
\]
Two documents collide if they share at least one band key. The only index-layer error is a
fixed-width hash collision: two distinct bands mapping to the same 64-bit key.
With $U$ distinct historical keys per band and $B$ bands, the per-document
residual is bounded by the union bound at approximately $B\,U/2^{64}$, which
is about $8.7\times10^{-10}$ at $U{=}10^{9}$ and
$8.7\times10^{-7}$ at $U{=}10^{12}$. This could be suppressed by working instead with 128-bit keys by another factor of $2^{64}$. Under the standard
MinHash-LSH analysis~\citep{indyk1998approximate,gionis1999similarity},
documents with shingle-set Jaccard similarity $J$ collide with probability
\[
    P_{\mathrm{LSH}}(J)=1-(1-J^R)^B .
\]
Thus, the shingle definition, $p$, $B$, $R$, and the collision policy define
the fuzzy-deduplication rule. 

Now we detail the design decisions specific to \puffer{}. We target efficient and governable fuzzy deduplication at dataset-release granularity, matching the needs of a growing corpus. Each
document $d \in D_k$ carries a stable identity $\operatorname{id}(d)=(f_d,r_d)$,
where $f_d$ is the source file and $r_d$ is the row within that file. Stable
identities are ordered lexicographically and are used for deterministic
within-dataset tie-breaking. Fig.~\ref{fig:index_structure} shows the
lifecycle of the fuzzy index for one LSH band; the same structure applies to all $B$ bands. Historical state for band $b$ is represented
as a set of immutable sorted segments,
\[
    \mathcal{H}_b=\{H_{b,1},\ldots,H_{b,L_b}\}.
\]
Each segment stores a duplicate-free sorted array of 64-bit band keys and is
accessed by memory mapping. The manifest records each segment's dataset tag,
level, live status, and, after compaction, the lineage information needed for
withdrawal (the set of origin datasets). Four operations act on this state: screen, commit, compact, and
withdraw. We describe each in turn.

Screening determines which documents from $D_k$ survive, before any new state
from the current dataset becomes query-visible. Within the current dataset,
\puffer{} groups documents by band key: for band $b$ and key $h$, the collision
group is $G_{b,h}=\{d\in D_k:v_b(d)=h\}$. \puffer{} retains the document with the smallest stable identity and marks the
remaining members for removal producing the internally deduplicated set $D'_k$. Bands are processed independently, and the
method deliberately does not construct a transitive duplicate graph across
all band collisions; this keeps the operation memory-bounded and avoids the
very large connected components that common boilerplate would otherwise
induce. We save a separate sidecar copy of this internally deduplicated dataset's band keys - this allows us to later withdraw from the index without complete rebuilds. Historical screening then checks each current band key against the
live historical segments for that band. During processing of dataset $D'_k$,
segments tagged with $k$ are excluded from the query view (excluding an unnecessary comparison of the same data to allow for a simple idempotent retry on crash),
$\mathcal{H}^{(-k)}_b=\{H\in\mathcal{H}_b:\operatorname{tag}(H)\neq k\}$,
and historical membership is evaluated by binary search over the
memory-mapped sorted segments (stopping when a match is found):
\[
    q(d)
    =
    \bigvee_{b=1}^{B}
    \bigvee_{H\in\mathcal{H}^{(-k)}_b}
    \mathbf{1}[v_b(d)\in H].
\]
A document is removed as a historical near-duplicate when $q(d)=1$. The
screened output $D''_k$ contains the documents that survive both
within-dataset and historical fuzzy deduplication.

Commit makes the screened dataset's index visible to future ingestion runs. For each
band $b$, \puffer{} sorts the band keys of the surviving
documents in $D'_k$ (not $D''_k$ the cross deduplicated set),
\[
    U_{k,b}
    =
    \operatorname{unique}
    \left(
    \operatorname{sort}
    \{v_b(d):d\in D'_k\}
    \right),
\]
and writes the array $U_{k,b}$ as an immutable sorted segment tagged with
dataset tag $k$. Note that we could instead retain only the keys corresponding to $D''_k$,
the fully deduplicated output. However, keys from later datasets that were
removed as historical duplicates would then be absent from the index. If the
dataset that originally caused their removal were subsequently withdrawn,
recovering the correct index state would require replaying later datasets or
rebuilding historical state. \puffer{} therefore retains the keys from $D'_k$,
including those removed only by historical screening, so withdrawal can be
performed through segment-local reconstruction without replaying subsequent
datasets. Segment files and manifest entries are written through
temporary paths and installed by atomic rename, so a segment becomes
query-visible only after the manifest update succeeds. Commit is
tag-replacing rather than blindly append-only: if a segment for tag $k$
already exists, the successful commit replaces the previous logical
contribution for that tag.

Compact bounds screening fanout by merging $T$ eligible same-level
segments within a band into one segment at the next level. The most
recently committed segment is withheld to preserve the simple retry
path. Larger $T$ reduces merge frequency but leaves more segments to
search, while smaller $T$ merges more eagerly and rewrites keys more
often. For input segments $H_1,\ldots,H_T$, compaction writes
\[
    H_{\mathrm{new}}
    =
    \operatorname{unique}
    \left(
    \operatorname{merge}[H_1,\ldots,H_T]
    \right).
\]
Because the inputs are sorted, \puffer{} performs this operation as a
streaming $T$-way merge under a fixed working-memory budget. Each key
is rewritten at most $O(\log_T K)$ times over $K$ releases, and the
merged segment contains exactly the union of the input keys.
Compaction therefore changes maintenance and screening costs, but not
membership decisions. It may, however, mix dataset contributions,
making subsequent withdrawal more expensive.

Withdrawal removes a dataset's contribution from the fuzzy index. Given a
victim dataset tag $j$, uncompacted or protected segments associated with
$j$ are removed directly from the live manifest, a constant-time metadata
operation. After compaction, the dataset's keys may be mixed with those of
other datasets in a merged segment. In this case, \puffer{} uses the saved
sidecar band-key state to reconstruct the affected segment from the
constituent datasets that contributed to it, excluding the withdrawn dataset.
The cost is therefore local to the affected segment, up to $\mathcal{O}(N \log K)$ if
the dataset has reached the base tier, but never requires rebuilding the
full historical index. Datasets with foreseeable withdrawal exposure can
instead be marked protected at ingest, excluding them from compaction and
preserving the constant-time withdrawal path at the cost of one additional
unit of query fanout each.

Retry refers to restarting the ingestion of a dataset after the pipeline is
interrupted or fails at any point during processing. \puffer{} makes such
restarts deterministic through stable dataset tags and manifest-atomic
commit. When processing $D_k$, segments tagged $k$ are excluded from the
historical query view, so a restarted run does not treat state previously
written for the same dataset as historical duplicates. All writes remain
attempt-local until commit: temporary segment files are not query-visible
unless referenced by the committed manifest, and a successful commit
replaces the logical contribution associated with tag $k$ rather than
appending a second copy. Consequently, under an unchanged historical index,
re-running the same dataset with the same configuration after an interruption
produces the same surviving documents and query-visible index state as an
uninterrupted run.

\section{Results}
\label{sec:results}

\begin{figure*}[t]
\centering
\begin{subfigure}[t]{0.45\textwidth}
\centering
\includegraphics[width=\linewidth]{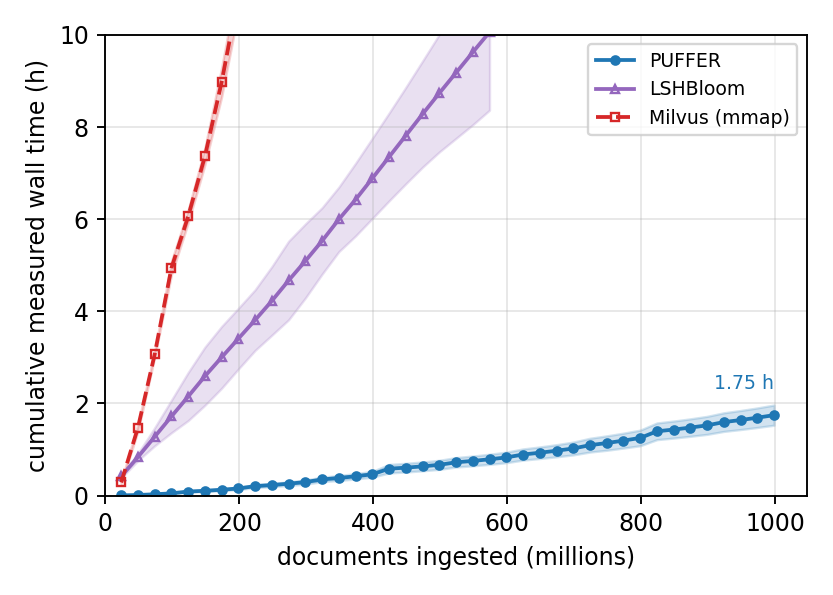}
\caption{Incremental ingestion.}
\label{fig:main-ingest}
\end{subfigure}
\hfill
\begin{subfigure}[t]{0.45\textwidth}
\centering
\includegraphics[width=\linewidth]{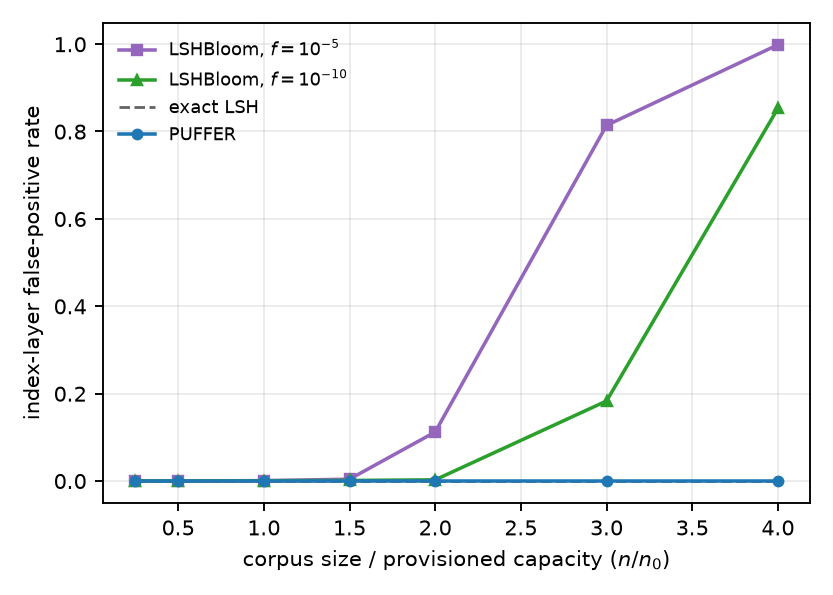}
\caption{Fidelity under corpus growth.}
\label{fig:main-accuracy}
\end{subfigure}

\medskip
\begin{subfigure}[t]{0.55\textwidth}
\centering
\includegraphics[width=\linewidth]{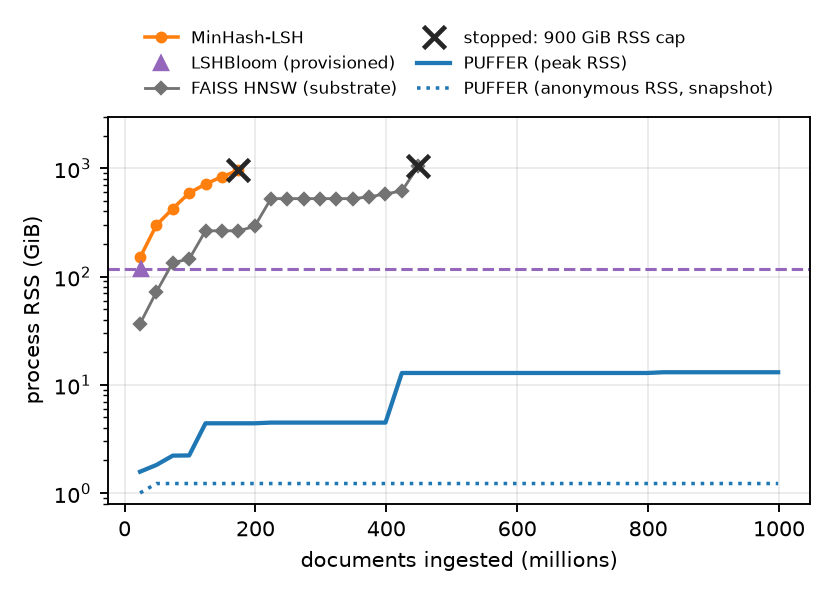}
\caption{Process memory during ingestion.}
\label{fig:main-memory}
\end{subfigure}
\caption{Comparison with the closest implemented alternatives. (a)
Cumulative measured wall time through one billion documents (up to ten hours) with $95\%$ CIs. Signature
preparation is excluded from the totals for all systems. The Milvus (mmap) series is client-observed cumulative wall time for the served MinHash-LSH baseline; within the ten-hour window it advances only to 187M documents. The LSHBloom series is the index-layer timing only, and reaches 600M documents in the ten-hour window. (b) False-positive rate on a known-unique holdout as
the corpus exceeds LSHBloom's provisioned capacity $n_0$ at two choices of $f_p$. (c) Process RSS
during ingestion; X marks runs stopped after exceeding a 900\,GiB RSS cap,
full MinHash-LSH at 956\,GiB after 173.7M documents and the FAISS HNSW
substrate (a stand-in for FOLD's memory scaling) at 1{,}043\,GiB after 449M vectors. The LSHBloom line corresponds to a provisioned scale of $n_0=2.5$ billion documents with $f_p=10^{-5}$. The dotted \puffer{}
line is an end-of-release private-memory snapshot rather than a peak. Full
measurement definitions are given in
Appendix~\ref{app:measurement}.}
\label{fig:main}
\end{figure*}

The central question in our evaluation is whether \puffer{} remains practical
as the historical corpus grows: can it preserve the configured MinHash-LSH
decisions while ingesting new releases efficiently, keeping memory demands
manageable, controlling maintenance cost, and supporting dataset-lifecycle
operations. We evaluate \puffer{} against these lifecycle requirements and compare it with the closest available MinHash-LSH baselines capable of repeated ingestion. In particular, the RAM wall limits the scalability of both exact MinHash-LSH and \fold{}~\citep{bore2026fold}. Because the \fold{} implementation has not been released as of the writing of this paper, we evaluate its memory scaling through a FAISS HNSW substrate~\citep{douze2024faiss} configured to match its reported index storage characteristics. \lshbloom{}~\citep{khan2024lshbloom} avoids this RAM wall through a fixed, upfront-provisioned Bloom filter, but its false-positive rate increases once the provisioned capacity is exceeded. Finally, in our throughput comparison against the implemented baselines, including Milvus, \puffer{} achieves the highest performance by a substantial margin. These results are summarized in Fig.~\ref{fig:main}.

These comparisons are performed at the index layer: we assume that the band keys have already been computed and evaluate the performance of duplicate identification and index construction. Accordingly, timings for document shingling and band-key computation are excluded. The index-layer
experiments feed each system the
same deterministic release stream of synthetic 64-bit band keys, or signatures
derived from them, with signature preparation excluded from timed phases. This allows us to verify the deduplication decisions of an exact MinHash-LSH oracle without incurring the $\mathcal{O}(N^2)$ cost of pairwise comparison. For the fidelity experiments, we control the distribution of Jaccard similarity among the underlying document representations. Near matches are fixed to Jaccard similarity $.9$ and non-matches are fixed at $J=0$. For the remaining experiments, we construct band key sets by sampling $30\%$ of the keys from a shared pool, with the remaining guaranteed to be novel. 
Technical details are provided in Appendix~\ref{app:measurement}. 


Baselines use released libraries with pinned versions, using public interfaces where the benchmark path permits: datasketch~1.10.0~\citep{datasketch_lib,khan2024lshbloom}, Milvus3.0-beta~\citep{milvus_minhash_lsh}, and FAISS~\citep{douze2024faiss}. In all tests, unless otherwise noted, we use $T=4$ for \puffer{}, which we show provides the best performance tradeoff for our experimental setting. Runs use one process per system, with no explicit
application-level parallelism with the exception of Milvus whose external server uses internal parallelism.
All runs use one node per timed job over a shared storage pool,
with per-release timing, memory, and decision digests checkpointed for
cross-system comparison. Timing and memory measurement notes as well as configuration details of the baseline comparisons are given in Appendix~\ref{app:measurement}.

\begin{figure*}[ht!]
\centering
\begin{subfigure}[t]{0.49\textwidth}
\centering
\includegraphics[width=\linewidth]{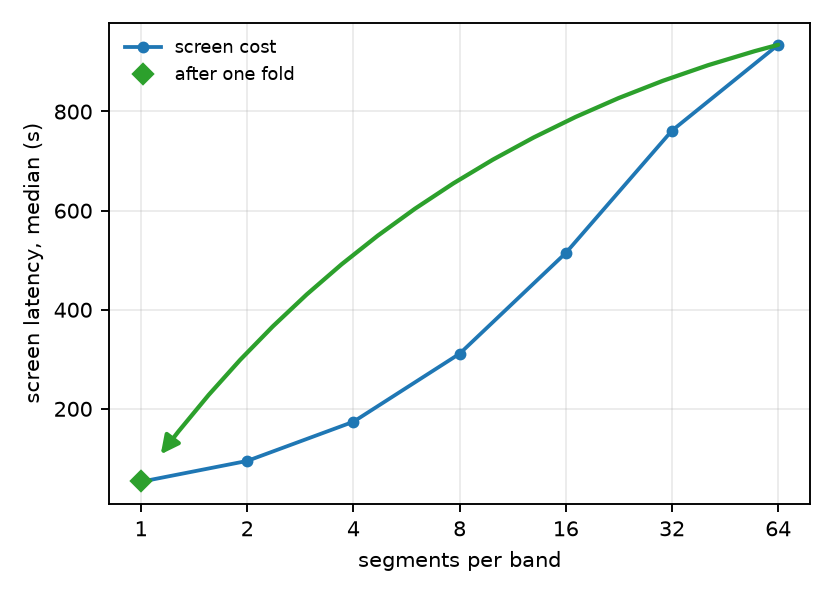}
\caption{Screening latency versus live segment count.}
\label{fig:dials-screen}
\end{subfigure}
\hfill
\begin{subfigure}[t]{0.49\textwidth}
\centering
\includegraphics[width=\linewidth]{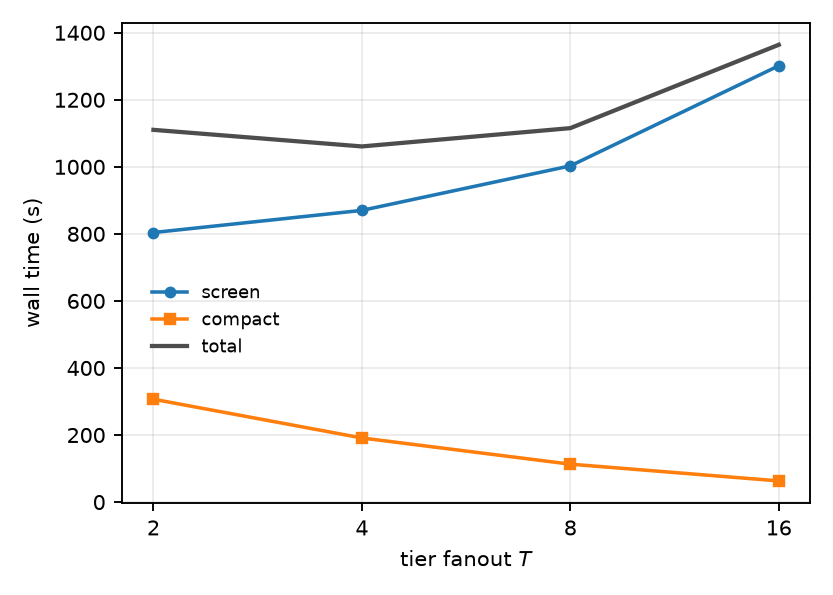}
\caption{Lifecycle cost versus tier fanout $T$.}
\label{fig:dials-fanout}
\end{subfigure}
\caption{Compaction is the performance dial. (a) Median latency to screen a
   release-sized batch against a fixed two-billion-key band divided into $S$
   segments; a 67-second streaming fold merges all 64 segments into one and
   restores single-segment latency. (b) Raising tier fanout $T$ defers merging,
   lowering compaction time while screening grows; their sum is minimized at
   $T{=}4$ for our system, although the optimum generally depends on factors such as the machine-specific relative costs of reading and writing data.}
\label{fig:dials}
\end{figure*}

Fig.~\ref{fig:main} asks whether our targeted properties continue to hold as the index grows,
examining ingestion speed, decision fidelity, and resident-memory demand.
Together, these measurements capture the main tension in large-scale
incremental deduplication: maintaining an ever-growing history without making
each new release progressively impractical to process.

Fundamentally, speed determines how practical a pipeline is for large ingestion scaling. Fig.~\ref{fig:main-ingest} shows incremental ingestion throughput: on the same deterministic sequence of input releases used for all systems, \puffer{} completed one billion documents across forty releases at
$B{=}16$ bands, with every screen, commit, and compaction included, in 1.75
hours of index time in a single process. For equal-sized releases, the quasilinear cumulative cost contrasts with the $\Theta(K N)=\Theta(N^2)$ cost of repeated rebuilding and is enabled by tiered compaction; Appendix~\ref{app:complexity} provides the full complexity derivation. The tradeoff is write amplification: in this experiment, compaction incurs 1.36 additional writes per key, with keys reaching the highest level rewritten up to $\log_T(K-1)$ times. Balancing this maintenance cost against reduced screening time allows overall throughput to be optimized. In the ten-hour throughput comparison, \puffer{} achieved
a $35\times$ speedup over the served MinHash-LSH baseline
(Milvus 3.0-beta, our configuration). This matters operationally because an incremental index is
useful only if new releases can be incorporated without historical state
becoming a growing ingestion bottleneck; in our comparison, \puffer{} sustains
billion-document ingestion while the served MinHash-LSH baseline did not reach the same
corpus scale within the benchmark window.

Next, we compare the fidelity of \lshbloom{} and \puffer{} when comparing deduplication decisions to a MinHash-LSH oracle. Fig.~\ref{fig:main-accuracy} shows the capacity effect: with genuine MinHash signatures over controlled
near-duplicate structure, \puffer{} tracks the exact-LSH oracle's answers at zero
index-layer false positives at every corpus size, with no capacity parameter
to set. \lshbloom's error grows once its provisioned capacity is exceeded because
the capacity must be chosen in advance. In particular, when the final corpus size is not known in advance, \puffer{} provides a practical alternative that does not require upfront capacity provisioning.

The memory comparison in Fig.~\ref{fig:main-memory} uses the same ingestion workload across systems. On disk, \puffer{} requires 8 bytes per band key, or about 128 bytes per document at $B{=}16$. Supporting dataset withdrawal without a full rebuild doubles this storage requirement. Even then, the index remains roughly $25\times$ smaller than the classical resident MinHash-LSH table.
At ten billion ingested keys the final footprint was 55.1\,GiB per band
under the tiered policy against 62.3\,GiB under the legacy selector, whose
frozen shards also permanently retain 18 percent duplicate keys they can
never merge away. Resident memory is elective, since screening touches
segments through memory mapping and the page cache: \puffer's process high
water at one billion documents is 16.3\,GiB, a peak that includes elective
mmap file-backed residency, while its end-of-release private-memory
snapshots stay near 1.2\,GiB. The
datasketch full MinHash-LSH baseline, by contrast, grows a mandatory
resident structure at roughly 6.5\,KB per document, and 956\,GiB at 173.7M documents, where it
exceeded a 900\,GiB RSS cap and stopped. To approximate the memory scaling of \fold{}, we use FAISS HNSW configured to match its reported index characteristics, since both maintain a resident proximity graph. At graph degree $M{=}128$
over $4096$-bit signatures, matching the configuration reported by \fold{}, the process measured ${\approx}1.6$\,KB per vector and
grew to $1{,}043$\,GiB at $448.7$M vectors, where it too crossed the
$900$\,GiB cap and stopped at release $17$ out of the planned $40$. Because the
graph must stay resident for traversal, this memory footprint cannot be made optional through memory mapping or reduced through compaction; like the frozen legacy
shards above, the graph is retained in full and only grows, so this substrate
reaches the resident-memory ceiling well short of the billion-document scale
\puffer{} sustains. Taken together, these results show that, among the approaches evaluated, \puffer{} is the only one that simultaneously accommodates the scalability and lifecycle requirements of a continuously growing corpus.

Immutable segments create a simple tradeoff: leaving more segments separate
avoids maintenance writes but makes every future screening operation search
more files, while merging them reduces screening cost at the price of rewriting
keys. Fig.~\ref{fig:dials} quantifies this tradeoff at production scale, using approximately 2 and 2.4 billion keys per band in parts (a) and (b), respectively. 
Fig.~\ref{fig:dials-screen} explicitly demonstrates how compaction provides a practical control over this
tradeoff by controlling the number of live segments. The read-side price of fanout is near-linear in segment count, as the
$S\cdot\mathcal{O}(\log(N/S))$ analysis predicts, and one streaming merge
restores
single-segment latency: a 67-second merge at the two-billion-key scale recovers
a 17.8$\times$ screening-latency reduction without changing membership
decisions. However, naive iterated compaction is itself not the answer as each compaction is a merge of sorted lists which also has a cost scaling linearly with the input lengths and cumulatively this would generate a quadratic ingestion cost for equal-sized releases. By using tiered merging at a fixed-fanout $T$, each key at worst is merged $\log_TK$ times - which is also the scale of the number of segments. As shown in Appendix~\ref{app:complexity}, this segment count combined with the cost of screening leads to an asymptotic cost to ingest all $N$ keys of $\mathcal{O}(N \log N \log_T K)$. This then gives us $T$ as a control over the trade-offs, raising $T$ defers merging, which
cuts write amplification from 2.64 at $T=2$ to 0.73 at $T=16$, at the cost of
screening over more segments. In Fig.~\ref{fig:dials-fanout}, we evaluate multiple values of $T$, making the screening--compaction trade-off explicit and identifying $T{=4}$ as optimal for this set of runs. This optimum is machine-dependent and may vary with operating conditions; it should not be interpreted as universal. The fanout therefore acts as a direct performance knob rather than changing
which documents are identified as duplicates.

We model the read and write costs of a large (but realistic) stream of $K$ ingestions of $m$ documents in Appendix~\ref{app:lambert}. In this limit, we are motivated to approximate the stream of ingested datasets as containing a fixed fraction $\nu$ of novel documents, whereas the rest overlap with those seen before. We then also assume both that each overlapping key has a flat probability distribution of when it was first seen (at time of its next ingestion), and that duplicated keys are immediately removed (instead of at compaction time). With these approximations, the resulting model has enough analytical structure to efficiently evaluate the sum of read and write costs, and hence, the optimal $T$ if machine specific costs to reading and writing data are provided ($c_r$ and $c_w$). The model also admits a coarse closed-form continuous approximation for the optimal $T$, which improves as $\log_T K$ grows larger, and is valid while $\frac{\ln(K/T)}{2\ln(\nu m)}$ remains small:
\begin{equation}
    T_{\mathrm{Lambert}}=\exp\!\left[1+W_0\!\left(\frac{\rho-1}{e}\right)\right],
\end{equation}
where $\rho=2c_w/(c_r\log_2 (\nu m))$ and $W_0$ is the principle branch of the Lambert $W$ function. In the regime where the this approximation holds, $K$ does not appear in the expression for optimal $T$, it is only a function of the relative write to read speed $c_w/c_r$ and the novel keys added per dataset $\nu m$. As $W_0$ is monotonically increasing, $T$ monotonically increases as a function of $\rho$. This quantifies the intuition of the trade-off; as the ratio of write to read cost increases, the optimal $T$ increases to reduce write amplification. As the number of new novel elements $\nu m$ increases, the cost to screen segments increases, so the optimal $T$ decreases to reduce the screening cost with more frequent compactions.
Note that in our case $\log_T K \sim 3.3$ and $\frac{\ln(K/T)}{2\ln(\nu m)}\sim .1$ so the approximation entails a non-negligible error (we also provide a more accurate but less transparent approximation in terms of generalized Lambert functions). Fortunately, the approximation's primary job is identifying adjacent integer values of $T$ as candidates for the optimum. Applied to our case with read and write coefficients measured from regression based on the the experiments we obtain $T_{\mathrm{lambert}}\approx 3.8$ which indicates (correctly) that we should expect an optimal $T$ of $3$ or $4$. A small caveat; when considering the exact sum of costs, it can be observed that the true optimal $T$ oscillates primarily between the two nearest integers to $T_{\mathrm{Lambert}}$ as $K$ grows due to cycles that track the digit decompositions of $K$ in various bases (compactions near ends of run are costs that are never amortized over future screenings).

We also evaluate a production-scale compaction run and confirm the same tradeoff at
ten billion ingested keys per band. With the default fanout $T=4$,
the tiered selector schedules due promotions and sends merges that
exceed the in-memory budget to the streaming executor. The completed
run has compaction write amplification of 2.42 writes per key while
preserving the same key unions and per-release decisions. 
Larger merges are executed through the memory-bounded streaming path rather than materialized fully in RAM.

Finally, we test whether the lifecycle guarantees that motivate \puffer{} remain
practical after the index has grown and undergone compaction. Table~\ref{tab:withdrawal-scale} withdraws one release from the completed
ten-billion-key index using the implemented withdrawal procedure, choosing victims
at each compaction depth. An uncompacted release withdraws in 2
milliseconds, a metadata-only operation whose cost is independent of corpus
size. After compaction the cost is level-local and proportional to the
affected segment, reaching 353 seconds for a victim merged into a
4.6-billion-key tier. The affected segment is reconstructed from the retained band-key sets of its
constituent datasets, excluding the withdrawn dataset, and post-withdrawal
validation found zero victim-exclusive keys remaining in every case. \puffer{} also offers protected tags if a dataset is understood to have a high likelihood of requiring withdrawal later which would maintain the uncompacted withdrawal speed.

Crash-retry determinism is verified separately: runs killed mid-commit at
every shard boundary and then retried produce bit-identical per-band state
under SHA-256 comparison, and identical decisions. These experiments show that lifecycle operations do not require discarding
and rebuilding the accumulated index: withdrawal remains local to the affected
state, while interrupted ingestion can be restarted without changing the
resulting index.

Beyond these controlled index-layer experiments,
Appendix~\ref{app:end_to_end} demonstrates end-to-end operation on
approximately 2.5 billion real documents partitioned into 100 Parquet
releases, completing incremental deduplication, cumulative index construction,
and cleaned-Parquet output in approximately five hours using eight nodes;
the per-release duplicate-removal rate increases as historical coverage
accumulates, as expected for cross-release deduplication.

\begin{table}[H]
\centering\footnotesize
\setlength{\tabcolsep}{3pt}
\caption{Withdrawal from a ten-billion-key band using our implementation. Index-side wall time includes planning, streaming reconstruction of the affected segment, and manifest update.}
\label{tab:withdrawal-scale}
\begin{tabular}{lrrr}
\toprule
victim state & affected keys & index-side wall  \\
\midrule
uncompacted ($L0$) & $2.5\times10^{7}$ & $0.002$\,s  \\
tier 1 & $9.7\times10^{7}$ & $3.5$\,s  \\
tier 3 & $1.2\times10^{9}$ & $83.7$\,s \\
tier 4 (base) & $4.6\times10^{9}$ & $353$\,s \\
\bottomrule
\end{tabular}
\end{table}

Taken together, these results show that \puffer{} does not obtain scalability
by sacrificing fidelity, memory efficiency, or dataset-level control. Among
the systems we evaluate, it is the only design that simultaneously preserves
the configured MinHash-LSH membership decisions as the corpus grows, avoids
a corpus-size-proportional resident index, provides an explicit control over
the screening--maintenance tradeoff, and supports deterministic retry and
dataset-scoped withdrawal without corpus-wide rebuilding. More broadly, these
properties make it practical to treat deduplication state as persistent
infrastructure for a continuously evolving training corpus, rather than as
temporary state that must be reconstructed whenever new data arrive or old
data must be removed.

\section{Discussion}
\label{sec:discussion}


\puffer{} is designed around a specific systems observation: for corpora
that evolve through repeated releases, fuzzy-deduplication state is more
useful as persistent infrastructure than as an artifact rebuilt for each
snapshot. Our results show that MinHash-LSH band-key state can be
maintained incrementally without a corpus-size-proportional resident
index, while preserving the configured membership rule and supporting
dataset-level retry and withdrawal. This problem may become increasingly
relevant beyond pretraining: post-training corpora contain prompts,
responses, reasoning traces, tool-use trajectories, and synthetic
rollouts for which the appropriate unit or rule of duplication may
differ. \puffer{} addresses the index-maintenance problem for lexical
MinHash-LSH; extending the same lifecycle principles to other
representations is a natural direction for future work.

A persistent segmented index is not necessarily the best design for
every workload. Snapshot-oriented pipelines remain attractive when
corpora are rebuilt infrequently, conventional in-memory MinHash-LSH is
simpler when the historical state fits comfortably in memory, and
Bloom-filter approaches such as LSHBloom are compact when corpus
capacity can be provisioned and controlled false positives are
acceptable. Database-backed or graph-based approaches such as Milvus
and FOLD instead favor served search, online admission, or approximate
neighbor retrieval. \puffer{} targets the regime of very large corpora
arriving as successive releases, where historical state must persist
without resident memory growing proportionally with corpus size and
where dataset-level retry, replacement, or withdrawal are required. However, we have also verified through our experimental comparisons (throughput and memory use) that compared to these baselines \puffer{} out performs conventional MinHash-LSH (as served by Milvus) and \lshbloom{}, and that this advantage is present at first build. 

The segmented layout also makes compaction an explicit optimization
problem. Increasing fanout $T$ reduces write amplification but leaves
more segments for future screening; decreasing $T$ has the opposite
effect. Under the workload model in Appendix~\ref{app:lambert}, these
costs can be modeled. In that model, many assumptions like fixed dataset novelty and flat distributions on the historical index where a duplicate key finds its pair allowed for a reasonable approximation. However, this model may not be appropriate for situations where the overlap between datasets depends on time. It is an interesting open direction for future research to model the temporal correlations of dataset overlap in different settings and reconsider the optimal fanout. Although $T=4$ was optimal in our
experiments, the appropriate value depends on release overlap and the
relative costs of reading and writing. A similar tradeoff governs
dataset lifecycle operations: protected or uncompacted datasets retain
constant-time withdrawal, while withdrawal after compaction requires
reconstructing the affected segment. Protection therefore exchanges
withdrawal cost for persistent query fanout and should be reserved for
datasets with plausible replacement or withdrawal risk.

An important distinction appears between maintaining the
\emph{deduplication index} and maintaining the materialized
\emph{deduplicated corpus} under withdrawal. We have shown that it is possible to efficiently
and faithfully remove the impact of a dataset from the index's
state needed for future screening without rebuilding the full
historical index, but it does not automatically restore documents
previously rejected because of that dataset. Reconstructing the
corresponding corpus as if the removed dataset had never been added requires replaying affected
releases. This counterfactual decision set could be evaluated for datasets that are expected to be removed and applied for a quick corpus rebuild as well, however, this is costly in terms of storage and screening. Much of the difficulty related to dataset withdrawal (and our proposed solution) is a result of \puffer{}'s first-seen incremental
policy, so representative selection may depend on release order and,
under the current policy, release partitioning. Applications requiring
an order-independent canonical corpus therefore need an additional
global representative-selection policy and would be interesting to consider generalizations of our methodologies in these settings.

Finally, exact preservation of the index rule should not be confused
with a unique definition of duplication. Text normalization, document
boundaries, shingle construction, MinHash length, and banding determine
which similarities the system can detect; \puffer{} assumes these choices
remain fixed for an index generation. The 64-bit band keys also retain
the residual collision probability quantified in
Section~\ref{sec:methods}, and lexical MinHash-LSH remains complementary
to semantic deduplication. Most controlled experiments in
Section~\ref{sec:results} begin from precomputed band keys, while
Appendix~\ref{app:end_to_end} demonstrates the complete path from real
Parquet input to cleaned output. Absolute screening performance depends
on storage and page-cache behavior, and the current implementation
serializes index commits rather than providing a distributed
transactional index.

\bibliographystyle{tmlr}
\bibliography{puffer_references}

\clearpage
\appendices

\section{Related work}
\label{app:related}

Fuzzy deduplication is commonly built on document resemblance, MinHash, and
locality-sensitive hashing~\citep{
broder1997resemblance,
indyk1998approximate,
gionis1999similarity,
leskovec2020mining}. Existing corpus-curation toolkits are largely organized
as batch workflows over supplied input collections. Dolma's public toolkit
documents Bloom-filter-based document and paragraph deduplication, while
DataTrove and NeMo Curator provide MinHash-LSH-style fuzzy deduplication
pipelines over fixed candidate pools~\citep{soldaini2024dolma,penedo2024datatrove,nemo_curator}.
These workflows may persist intermediate artifacts such as signatures, buckets,
duplicate identifiers, or filtered outputs, but their public interfaces do not
expose a persistent, queryable MinHash-LSH state that can be incrementally
updated, retried, compacted, or withdrawn at dataset granularity. Consequently,
adding a new release generally requires re-running global candidate-generation,
matching, clustering, or filtering stages over the accumulated release artifacts,
rather than querying and updating a maintained historical index.

Embedding-based methods such as SemDeDup and D4 target semantic redundancy
and data diversification~\citep{abbas2023semdedup,tirumala2023d4}. That
question is complementary to the index-layer problem studied here: once a
deduplication rule has been selected, the historical index should preserve that
rule as the corpus evolves. Although our implementation uses MinHash-LSH, the index-layer question is conceptually separate from research on the underlying duplicate-detection rule. The representation may also generalize to rules that can be expressed as fixed-width membership keys.

Recent systems have advanced the index layer in different directions.
\lshbloom{} compresses MinHash-LSH state by representing band membership
with Bloom filters, reducing index size at the cost of probabilistic false
positives whose rate is fixed by a provisioned
capacity~\citep{khan2024lshbloom,bloom1970spacetime,almeida2007scalable}.
Bloom filter state is also not invertible, so dataset withdrawal requires
rebuilding from surviving sources. Milvus/Zilliz exposes MinHash-LSH as a
database-level indexing capability for approximate deduplication and
similarity search over MinHash signatures~\citep{milvus_minhash_lsh}; the
index is served, and in our measurements of throughput we found a $35\times$ speedup over this baseline on ingesting one billion documents. \fold{} targets online fuzzy
deduplication by maintaining an approximate-nearest-neighbor index over
admitted documents using graph-based retrieval~\citep{bore2026fold,malkov2020hnsw}. It addresses the repeated global-candidate-generation problem through online
retrieval, reports 94 to 97 percent recall relative to a baseline approximation on corpora up to 30M documents, holds its
graph resident (a process-RSS-delta estimate around index construction on a
FAISS binary-HNSW substrate~\citep{douze2024faiss} at its reported degree
gave roughly 1.2 to 1.7\,KB per document), and
does not report a dataset-withdrawal operation. This causes \fold{} to encounter a resident-memory wall as the corpus grows, while providing lower fidelity to the MinHash-LSH oracle.

\puffer{} occupies a distinct point. Unlike snapshot pipelines, it treats
corpus construction as ongoing maintenance. Unlike Bloom-filter membership,
it stores materialized band keys in sorted memory-mapped segments, enabling dataset withdrawal without fully rebuilding the index, and avoiding
index-level false positives beyond the fixed 64-bit collision residual. Unlike online graph approaches, it
targets reproducible dataset-level batch ingestion rather than low-latency
document-by-document admission. Unlike database-integrated MinHash-LSH
search, it makes dataset lifecycle control a first-class property of the
index itself, with provenance carried by the physical layout. The storage
machinery draws directly on LSM-tree compaction
theory~\citep{oneil1996lsm,luo2020lsmsurvey,dayan2018dostoevsky,dong2021rocksdb},
but dataset tags make the deletion unit explicit in the layout.

\section{Complexity Derivation}
\label{app:complexity}
 
All bounds in the following derivation are costs per band. We consider the operating costs of a fixed-fanout tiered compaction based ingestion. Throughout, $K$ denotes the number of
committed releases, each of which initially contributes one level-0 ($L0$) segment per band; $m_k$ the number of band keys (per band)
contributed by release $k$, $m_{\max}=\max_{1 \le k \le K} m_k$, $N=\sum_{k=1}^{K} m_k$ the total
number of band keys ingested over the corpus lifetime, $T\ge2$ the
max per tier fanout, and $P$ the number of protected $L0$ segments, which are excluded from compaction. $T$, and $P$ are configuration constants;
$K$ and $N$ are unbounded. Throughout this appendix, $\log$ without an explicit base denotes $\log_2$. 
 
\textit{Live-segment count}:
By induction on the merge rule ($T$ same-level segments merge into one
segment at the next level), a level-$\ell$ segment is the union of
exactly $T^{\ell}$ $L0$ arrivals. With $K$ arrivals the maximum occupied
level is therefore $L\le\log_TK$. The selector re-merges any level that
accumulates $T$ unprotected segments, so at rest each of the $L{+}1$
levels holds at most $T{-}1$ of them, and the live-segment count obeys
\[
S \;\le\; (T-1)(L+1)+P
\;=\;\mathcal O\!\bigl(T\log_TK+P\bigr)
\;=\;\mathcal O(\log_T K),
\]
for fixed $T$, and $P$.

\textit{Screening}:
Each ingested key is binary-searched against every live segment. A
segment holds at most $N$ keys, so one probe costs $\mathcal O(\log N)$
comparisons, and a key costs at most
$S\cdot\mathcal O(\log N)=\mathcal O\bigl((T\log_TK+P)\log N\bigr)
=\mathcal O(\log N\log_T K)$. Over all $N$ ingested keys, screening is
$\mathcal O(N\log N\log_T K)$: the asymptotically dominant term.
 
\textit{Commit}:
Committing a release sorts and deduplicates its surviving keys once
before the append: $\mathcal O(m_k\log m_k)$ for release $k$. Summed
over releases,
$\sum_{k=1}^K m_k\log m_k\le N\log m_{\max}\leq N\log N)$. Therefore, commits contribute a total cost $\mathcal{O}(N \log N)$.
 
\textit{Compaction}:
Two factors multiply. (i)~Rewrite count: a key is rewritten only when its
containing segment's level increases, and levels are bounded by $L$, so each
key is rewritten at most $\mathcal{O}(\log_T K)$ times. (ii)~Per-rewrite
cost: during merge, writing each key is of $\Theta(1)$. 
Hence, the total compaction is
$\mathcal{O}(N\log_T K)$. 
 
\textit{Withdrawal}:
While the victim's segments are uncompacted or protected, withdrawal is a
manifest unlink, $\Theta(1)$ per band independent of corpus size.
After compaction it is the streamed merging of the set of sorted lists of keys from each of the surviving dataset that participated in the compacted segment, up to $\Theta(N \log K)$ when the victim
reached the base tier, level-local, never a whole-corpus rebuild.
 
\textit{Total}:
Screening at $\mathcal O(N\log N\log_T K)$ plus commit and compaction at
$\mathcal O(N\log N)$ give the cumulative bound:
\[
\mathcal O(N\log N\log_T K)
\]
holding for fixed fanout $T$, and a bounded protected set $P$. No term requires a cap on the number of releases
or on capacity provisioning; all constants are configuration constants.
For fixed release size, repeated snapshot rebuilding costs
$\Theta(KN)$ because each of the $K$ releases requires processing the
corpus accumulated up to that point. This comparator assumes the
optimistic case of linear work per rebuild; additional sorting,
candidate generation, or reindexing can only increase the cost.

\section{Measurement notes}
\label{app:measurement}

The following notes define the timing and memory quantities used in the
comparison figures.

\textit{Placement and hardware}: All timed runs execute on nodes of a single
hardware class; dual-socket Intel Xeon Platinum~8570 servers ($2$ sockets
$\times\,56$ physical cores $=112$ cores, $224$ logical CPUs under two-way SMT,
two NUMA domains), ${\approx}2$\,TiB DRAM, Ubuntu~22.04 (Linux~5.15). Unless a
metric explicitly states otherwise, each system is timed as a \emph{single
process}: the curves report per-process index-layer and algorithmic cost, so the
node's large core count bounds the available memory bandwidth and operating-system
page cache rather than being consumed by the measurement. Library-internal
parallelism is the only exception and is flagged where it occurs (the FAISS
substrate runs library-native OpenMP across all $224$ logical CPUs and makes no
deduplication-decision claim); \puffer's optional Ray-based multi-node executor is
out of scope for these comparison figures. Input release streams and all heavy
state (sorted band-key segments, Bloom and embedded-store files, intermediate
merge shards) reside on a \emph{shared} CephFS pool (multi-monitor Ceph cluster,
$4$\,MiB object stripe), not node-local disk. Shared storage reflects the
intended deployment but is also a confounder: cross-tenant metadata,
client-cache, and I/O contention can perturb absolute wall times, so within a run
we compare scaling \emph{shapes} and treat absolute constants cautiously across
runs.

\textit{Timing scope}: All systems consume the same deterministic release stream and
exclude signature preparation from timed phases. In Fig.~\ref{fig:main-ingest},
the LSHBloom checkpoints (one per 25M-document release) measure screen plus insert through the
benchmark's post-banding internal storage adapter at multiples of 25 million documents. This isolates Bloom lookup and storage after band-key generation;
it should therefore be interpreted as an index-layer
component timing, not as public-API or end-to-end pipeline timing. \puffer{} likewise consumes
precomputed band keys, its published production path also includes shingling and evaluation of MinHash-LSH band keys which we skip to compare the index building directly to other baselines. Under this harness, signature
preparation adds roughly 10{,}500 seconds per 25M-document release for the
datasketch baselines, so their full per-release wall is about 3.1 to 3.5
hours. The Milvus run (v3.0-beta, our configuration, 25M-document releases)
is plotted using the client-side wall-clock time for search operations
in each release. This is a lower bound on total release-completion time,
because server-side insertion, segment sealing, and index construction
proceed asynchronously and are not included in the measured interval. The FAISS series is a substrate resource reference at FOLD's
reported graph degree (128): it runs with library-native OpenMP parallelism (224
threads observed), makes no deduplication-decision claim, and its ingestion
timings are not comparable to the single-process curves.

\textit{Memory metrics}: Memory curves report peak process RSS
(\texttt{ru\_maxrss} high-water) per release; the FAISS values are raw
process RSS including harness buffers. The dotted \puffer{} series is an
end-of-release snapshot of anonymous RSS, the private mappings (heap,
stack, and other non-file-backed pages); mapped index segments are file-backed and reclaimable by the
operating system, but the high-water metric does not decompose peak
composition. LSHBloom's residency is set by its provisioned capacity (2.5
billion keys per band here) and measured after its first release.

\textit{Throughput/Memory experimental configurations}: All systems tested consume the same
deterministic release stream for the ingestion experiments: $m=25$M synthetic 64-bit band keys per release
($30\%$ resampled from a fixed $4m$-key pool as cross-release duplicates, the
rest novel). The throughput is averaged over three seeds. All are configured with $128$-permutation MinHash and $B=16$ bands.
PUFFER uses tiered fanout $T=4$ with a $4$\,GiB merge budget; LSHBloom
(datasketch~1.10.0) provisions Bloom capacity $n_0=2.5\times10^{9}$ at
$f_p=10^{-5}$; Milvus (v3.0-beta, MINHASH\_LSH) is operated in memory-mapped configuration with
server-side compaction every four releases.

\textit{Fidelity experimental configuration}: The inputs are generated from $128$-permutation MinHash over
$40$-shingle documents; a near-duplicate shares a calibrated subset of a base
document's shingles (Jaccard similarity $0.90$), while independently generated documents
draw disjoint shingle blocks and so have Jaccard similarity $0$ by construction. Bloom
capacity $n_0=10^{5}$ is overfilled by factors $0.25$ to $4$ at
$f_p\in\{10^{-5},10^{-10}\}$, with false positives measured on a 2,000-document
known-unique holdout and $200$ near-duplicate equivalence queries.

\begin{figure*}[ht!]
    \centering
    \includegraphics[width=.6\linewidth]{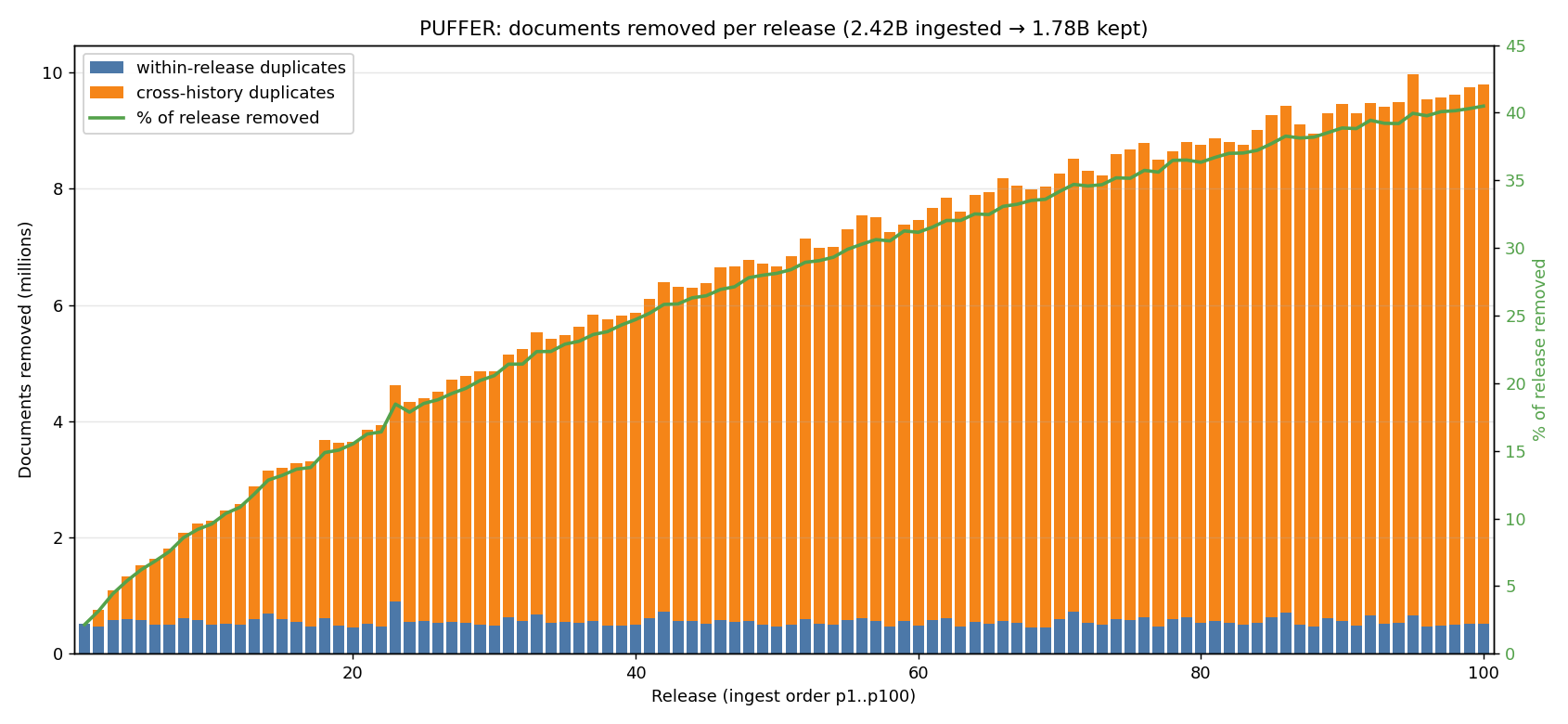}
    \caption{Document removal fraction during incremental fuzzy deduplication
    of 100 real Parquet releases derived from
    \texttt{allenai/dolma3\_pool}, totaling approximately 2.5 billion
    documents. Each release contains roughly 25 million documents. Within-release duplicate removal remains
approximately stable across releases, as expected for similarly sized
input partitions, while cross-release removal increases as each release
is screened against the growing historical index. Consequently, 
    the total fraction removed rises from approximately 2\% for the first release
    to approximately 40\% by the final releases.}
    \label{fig:real_parquet_dedup}
\end{figure*}

\section{END-TO-END INGESTION ON REAL PARQUET RELEASES}
\label{app:end_to_end}

To complement the index-layer experiments in Section~\ref{sec:results}, we
also evaluated \puffer{} on a large end-to-end ingestion workload using real
Parquet data derived from \texttt{allenai/dolma3\_pool}~\citep{olmo2025olmo3}. The workload contains
approximately 2.5 billion documents, partitioned into 100 releases of roughly
25 million documents each. Releases are processed sequentially, so that each
new release is deduplicated against the historical index accumulated from all
previous releases. The run begins from input Parquet files and produces both
the cleaned Parquet corpus and the cumulative fuzzy-deduplication index.

Fig.~\ref{fig:real_parquet_dedup} shows the documents removed
from each successive release, separated into within-release and cross-release duplicates. The horizontal axis denotes release order, while
the vertical axis reports the fraction of documents removed from that release
as duplicates against the accumulated history. Importantly, this is a
per-release removal fraction rather than a cumulative removal statistic: the
historical index grows along the horizontal axis, increasing the opportunity
for a newly arriving document to match content seen in earlier releases.

The two components in Fig.~\ref{fig:real_parquet_dedup} exhibit the expected
behavior of incremental deduplication. Within-release removal remains
approximately stable across the sequence: each release is of similar
size, and this component depends only on redundancy within the incoming
release rather than on the accumulated history. Cross-release removal,
in contrast, increases steadily as the historical index covers an
increasingly broad set of previously observed content. The probability
that a document in a new release matches earlier material therefore
grows with release order. Together, these effects raise the total
removal fraction from roughly 2\% in the first release to about 40\%
near the hundredth release, demonstrating both stable within-release
deduplication and increasing detection of redundancy across release
boundaries.

The complete workload finished in approximately five hours using eight
worker nodes. This timing includes processing the real Parquet releases,
incremental fuzzy-deduplication against the growing history, construction and
maintenance of the cumulative index, and writing the resulting cleaned
Parquet outputs. The experiment therefore complements the controlled
index-layer measurements by demonstrating that the same design can be used
as a practical end-to-end data-processing pipeline at multi-billion-document
scale.

\section{Optimal Fanout Modeling}
\label{app:lambert}

This appendix derives the fanout $T$ that minimizes the total
screening-plus-compaction work considered in
Appendix~\ref{app:complexity}. We study the case where overlapping keys
are removed at commit, so live segments partition the retained corpus,
and a duplicate probe stops at its first match. When duplicate keys are
retained until compaction this model acts only as an approximation. A
key element of this model is that it should be interpreted as the
cumulative costs of ingestion after a large corpus has already been
accumulated; in this limit a simple approximation is that there is a
fixed novelty fraction of each new set, whereas the rest of the
documents overlap with the previously accumulated corpus. Additionally,
we assume that there is no temporal correlation between appearances of
keys and that the probability of a key match is uniform throughout
releases.

\textit{Model}:
Each of $K$ releases presents $m$ keys, of which a fixed fraction $\nu$
is novel; write $o=1-\nu$, $q=\nu m$, and $M=K-1$ for the number of
screened releases, so $N=mK$ keys are ingested and $qK+om$ retained.
Conditional on being a duplicate, a presented key is uniformly
distributed over the retained keys. When release $k$ is screened, the
structure holds one protected segment (release $k-1$) and a base-$T$
counter over $k-2$: $d_\ell(k-2)$ (the base $T$, $\ell$th digit of
$k-2$) segments at level $\ell$, each holding $qT^{\ell}$ keys (the $om$
first-release offset does not affect any leading term and is
suppressed). The live-segment count is
$S_T(k)=1+\sum_\ell d_\ell(k-2)$. Segments are probed earliest-first,
which in the tiered layout is non-increasing size. Let $c_r$ be the cost
of one binary-search comparison and $c_w$ the cost of rewriting one
retained key during compaction.

\textit{Exact objective}:
Order the live segments $n_{k,1}\ge\cdots\ge n_{k,S_T(k)}$ and set
$c_{k,r}=\log_2n_{k,r}$, $M_{k,r}=\sum_{j\ge r}n_{k,j}$, and
$N_{k-1}=M_{k,1}$. First, we consider the screening cost. A novel key
scans everything, $F_T(k)=\sum_rc_{k,r}$; a duplicate reaches segment
$r$ exactly when its (unique) stored copy lies in segment $r$ or later,
so we can evaluate the expected duplicate key's cost $D_T(k)$ and the
cumulative expected screening comparison count $\Phi_K(T)$:
\begin{gather*}
D_T(k)=\frac1{N_{k-1}}\sum_{r=1}^{S_T(k)}M_{k,r}\,c_{k,r},\\
\Phi_K(T)=m\bigl(\nu\,\mathfrak F_K(T)+o\,\mathfrak D_K(T)\bigr),
\end{gather*}
where
\[
\mathfrak F_K(T)=\sum_{k=2}^{K}F_T(k),
\qquad
\mathfrak D_K(T)=\sum_{k=2}^{K}D_T(k)
\]
are the cumulative novel and duplicate comparison counts, kept separate
because they smooth differently below. Now we consider the write cost. A
carry into level $j$ occurs $\lfloor(K-1)/T^{j}\rfloor$ times and
rewrites $qT^{j}$ keys, so the total rewrite count is
\[
\Psi_K(T)
=q\sum_{j\ge1}T^{j}\Bigl\lfloor\frac{M}{T^{j}}\Bigr\rfloor.
\]
With the machine-specific read and write costs $c_r$ and $c_w$, the
optimization objective for the $T$ parameter is
\begin{equation}
\label{eq:exact-objective}
\begin{gathered}
\mathcal C_K(T)
=c_r\,\Phi_K(T)+c_w\,\Psi_K(T),\\
T_K^{\star}=\operatorname*{arg\,min}_{T\in\{2,\dots,K-1\}}\mathcal C_K(T).
\end{gathered}
\end{equation}
Direct enumeration of \eqref{eq:exact-objective} is the most reliable
final optimization, since it retains all floor terms, partial counter
cycles, and the integer restriction, and is easy to implement in code.
The remainder of this appendix shows that the novel and rewrite sums
admit closed digit forms, and produces smooth candidate generators.

\textit{Exact digit sums}:
A level-$\ell$ segment holds $qT^{\ell}$ keys, so each probe cost splits
into a base part and a level part,
$\log_2(qT^{\ell})=\log_2 q+\ell\log_2 T$, and summing over the live
segments gives the counter representation
\begin{gather*}
F_T(k)=S_T(k)\log_2 q+w_T(k-2)\log_2 T,\\
w_T(n)=\sum_{\ell\ge0}\ell\,d_\ell(n).
\end{gather*}
Both cumulative sums reduce to the per-level digit totals
$G_\ell=\sum_{n=0}^{M-1}d_\ell(n)$, which have a closed form: with
\[
\begin{gathered}
a_\ell=\Bigl\lfloor\frac{M}{T^{\ell+1}}\Bigr\rfloor,
\qquad
r_\ell=M-a_\ell T^{\ell+1},\\
b_\ell=\Bigl\lfloor\frac{r_\ell}{T^\ell}\Bigr\rfloor,
\qquad
c_\ell=r_\ell-b_\ell T^\ell,
\end{gathered}
\]
complete cycles of digit $\ell$ (period $T^{\ell+1}$), complete digit
blocks in the final partial cycle, and the final incomplete block give
\begin{equation}
\label{eq:digit-sum-exact}
G_\ell
=a_\ell T^{\ell+1}\frac{T-1}{2}
+T^\ell\frac{b_\ell(b_\ell-1)}{2}
+b_\ell c_\ell .
\end{equation}
Only finitely many $G_\ell$ are nonzero, so with
$\Sigma_0=M+\sum_\ell G_\ell$ (one protected segment per screened
release) and $\Sigma_1=\sum_\ell\ell\,G_\ell$,
\begin{equation}
\label{eq:novel-finite-exact}
\sum_{k=2}^{K}F_T(k)
=\Sigma_0\log_2 q+\Sigma_1\log_2 T ,
\end{equation}
evaluable in $\mathcal O(\log_TK)$ operations. The rewrite sum reduces
to the digits of $M$ itself: expanding
$\lfloor M/T^j\rfloor=\sum_{\ell\ge j}d_\ell(M)\,T^{\ell-j}$ shows that
$d_\ell(M)\,T^\ell$ appears once for each $j=1,\dots,\ell$, so
\begin{equation}
\label{eq:rewrite-digit-exact}
\Psi_K(T)=q\sum_{\ell\ge1}\ell\,d_\ell(M)\,T^\ell .
\end{equation}

\textit{Smooth approximation}:
Write $x=\ln T$, $A=\ln M$, and $L=\log_T M=A/x$. As $n=k-2$ sweeps
$0,\dots,M-1$, each digit equidistributes over $\{0,\dots,T-1\}$ across
complete cycles. For fixed $T$,
\begin{gather*}
\Sigma_0=M\left[1+\frac{T-1}{2}L\right]+\mathcal O_T(M),\\
\Sigma_1=M\frac{T-1}{4}L(L-1)+\mathcal O_T(ML),
\end{gather*}
and dropping the floors in the rewrite count gives
$\Psi_K(T)=qML+\mathcal O_T(qM)$. Thus the discarded partial-cycle and
floor terms are smaller than the retained terms by a relative
$\mathcal O_T(1/L)$, and
\begin{align}
c_r\nu m\sum_kF_T(k)
&\approx
c_rqM\Bigl[
\log_2 q
+L\tfrac{T-1}2\log_2 q \notag\\
&\qquad\qquad
+L(L-1)\tfrac{T-1}4\log_2 T
\Bigr], \\
c_w\Psi_K(T)&\approx c_wqM\,\frac{A}{x}.
\end{align}
For duplicates, approximate the counter by a state with the same
occupancy $\bar d$ at every level. Level masses are then
$\bar d\,qT^{\ell}$, so the top level holds a fraction
$\approx\tfrac{T-1}T$ of the stored keys \emph{independently of}
$\bar d$: occupancy sets how many segments each level contributes, not
the geometric mass ratio between levels. Under the uniform match
location and earliest-first probing, with probability $\tfrac{T-1}T$ the
match lies in the top level and, being uniform among its $\bar d$ equal
segments, is found after $\tfrac{\bar d+1}2$ probes on average; with
probability $\tfrac1T$ all $\bar d$ top segments are probed in vain and
the search recurses into the geometrically self-similar remainder. The
expected probe count $E$ therefore satisfies
\[
E=\frac{T-1}{T}\cdot\frac{\bar d+1}{2}+\frac1T\bigl[\bar d+E\bigr],
\]
that is, $E=\tfrac{\bar d+1}{2}+\tfrac{\bar d}{T-1}$, and the
trajectory-average occupancy $\bar d=\tfrac{T-1}2$ gives
$\bar E=\tfrac{T+3}4$. Probes concentrate on the largest segments, whose
sizes are within a constant factor of the corpus
$N_{k-1}\approx q(k-1)$, so each costs $\approx\log_2\bigl(q(k-1)\bigr)$;
the exact average over the $M$ screened releases is
\begin{align*}
\overline{B_M}
&=\log_2 q+\frac{\ln\Gamma(M+1)}{M\ln2}\\
&=\log_2(qM)-\frac1{\ln2}+\mathcal O(M^{-1}\ln M),
\end{align*}
so $c_rom\sum_kD_T(k)\approx c_rom\,M\bar E\,\overline{B_M}$.
Collecting the $T$-dependent terms and converting to natural logarithms,
\begin{equation}
\label{eq:csmooth}
\begin{aligned}
\mathcal C_{\mathrm{smooth}}(T)
=M\Bigl\{
&\gamma\,(T-1)\Bigl[\frac{H}{x}-A\Bigr]\\
&+\frac{B}{x}
+d\,(T+3)
\Bigr\},
\end{aligned}
\end{equation}
up to an additive constant, with
\[
\begin{gathered}
\gamma=\frac{c_rq}{4\ln2},
\qquad
H=A\bigl(A+2\ln q\bigr),\\
B=c_wqA,
\qquad
d=\frac{c_r\,om\,\overline{B_M}}{4}:
\end{gathered}
\]
in order, the level-dependent depth of the novel scans, the compaction
writes, and the early-stopping cost of duplicates.

\textit{Ordinary Lambert solution}:
We can split the novel bracket as
\[
\gamma(T-1)\Bigl[\frac Hx-A\Bigr]
=\frac{a(T-1)}{x}
+\gamma(T-1)\Bigl[\frac{A^2}{x}-A\Bigr],
\]
where $a=\tfrac12c_rq\log_2 q\,A$. The ratio of the level-dependent
remainder to the retained $a$-term is
\[
\frac{\ln(M/T)}{2\ln q}\approx0.10
\]
at the operating point. Additionally, a duplicate probes
$\bar E=\mathcal O(T)$ segments where a novel key scans all
$\Theta(TL)$, so the duplicate term is a roughly $10\%$ correction whose
variation is modest over the candidate range $T=2,\dots,5$. Dropping
both small terms leaves the bare tradeoff. Reads grow $\propto(T-1)$ as
levels widen, while writes fall $\propto1/\ln T$ as levels vanish. This gives us a simpler $\mathcal C$ to optimize over $T$ (differentiating by $T$ and setting to zero we obtain the equation for the optimal $T$):
\[
\widehat{\mathcal C}(T)=\frac Mx\bigl[a(T-1)+B\bigr],
\qquad
T(\ln T-1)=\rho-1,
\]
where $\rho=B/a=2c_w/(c_r\log_2 q)$. Hence
\[
T_{\mathrm{Lambert}}
=\exp\!\left[1+W_0\!\left(\frac{\rho-1}{e}\right)\right],
\]
where $W_0$ is the principle branch of the Lambert $W$ function. 
The factor $A$ cancels in $\rho$: the coarse optimum depends on the
workload only through $\log_2 q$, and on the machine only through
$c_w/c_r$ (notably the number of releases does not enter into this equation - it only appears through dropped terms).

\textit{Generalized-Lambert refinement}:
Retaining all of \eqref{eq:csmooth}, differentiating with
$\mathrm dx/\mathrm dT=1/T$, multiplying by $Tx^2$, and setting $T=e^x$
gives the scalar stationary equation
\begin{equation}
\label{eq:scalar}
e^{x}\Bigl[-(\gamma A-d)\,x^{2}+\gamma Hx-\gamma H\Bigr]
=B-\gamma H .
\end{equation}
If $\lambda=\gamma A-d=0$, this reduces to the ordinary-Lambert form
\[
T=\exp\!\left[1+W\!\left(\frac{B-\gamma H}{e\gamma H}\right)\right].
\]
For $\lambda\ne0$, factoring the quadratic yields
\begin{gather*}
e^{x}(x-r_{+})(x-r_{-})
=\frac{\gamma H-B}{\lambda},\\
r_{\pm}
=\frac{\gamma H\pm\sqrt{\gamma^{2}H^{2}-4\lambda\gamma H}}{2\lambda},
\end{gather*}
the defining equation of the two-parameter generalized Lambert function~\citep{mezo2017lambert}, so
\[
T_{\mathrm{gen}}
=\exp\!\left[\genW{r_{+}}{r_{-}}{\frac{\gamma H-B}{\lambda}}\right].
\]

\textit{Numerical operating point}:
At $K=96$, $m=25\times10^{6}$, $\nu=0.7$, $c_r=1.24\,\mathrm{ns}$,
$c_w=34\,\mathrm{ns}$ ($c_r$ and $c_w$ are estimated somewhat circularly by regression on the ingestion cycles varying over $T$ illustrated in Fig.~\ref{fig:dials-fanout}): $q=17.5\times10^{6}$,
$\overline{B_M}\approx29.24$, $\rho\approx2.279$,
$T_{\mathrm{Lambert}}\approx3.80$, and \eqref{eq:scalar} gives
$T_{\mathrm{gen}}\approx3.49$. As an illustration of
\eqref{eq:digit-sum-exact}, at $T=4$:
$(G_0,G_1,G_2,G_3)=(141,141,111,31)$, so $\Sigma_0=95+424=519$ and
$\Sigma_1=456$. Exact enumeration of \eqref{eq:exact-objective}, using
\eqref{eq:novel-finite-exact} and \eqref{eq:rewrite-digit-exact} for the
novel and rewrite sums and the exact segment states for the duplicate
sum, gives (in seconds)
\[
\setlength{\arraycolsep}{3pt}
\begin{array}{c|ccccc}
T & 2 & 3 & 4 & 5 & 6\\
\hline
\mathcal C_K(T) & 544.90 & 506.92 & \mathbf{475.34} & 488.26 & 495.00
\end{array}
\]
and hence $T_K^{\star}=4$. 

The continuous roots retain the leading complete-cycle terms; the exact
objective restores the partial cycles, write floors, duplicate states,
and the integer restriction. They therefore serve only as candidate
generators: evaluate \eqref{eq:exact-objective} at the integers
neighboring $T_{\mathrm{Lambert}}$ and $T_{\mathrm{gen}}$ and take the
minimum, or enumerate all $T\in\{2,\dots,K-1\}$ when this range is
small.

\end{document}